# Charge Transport in Single Au|Alkanedithiol|Au Junctions: Coordination Geometries and Conformational Degrees of Freedom

Chen Li[1], Ilya Pobelov[1], Thomas Wandlowski[1,2] [*], Alexei Bagrets[3], Andreas Arnold[3], and Ferdinand Evers[3,4] [*]

[1]Institute of Bio- and Nanosystems IBN 3 and Center of Nanoelectronic Systems for Information Technology, Research Center Jülich, D-52425 Jülich, Germany
[2] Departement of Chemistry and Biochemistry, University of Bern, CH-3012-Bern, Switzerland
[3]Institute of Nanotechnology, Research Center Karlsruhe, PO Box 3640, D-76021 Karlsruhe, Germany
[4]Institut für Theorie der Kondensierten Materie, Universität Karlsruhe, D-76128 Karlsruhe, Germany

**Abstract**

Recent STM molecular break-junction experiments have revealed *multiple* series of peaks in the conductance histograms of alkanedithiols. To resolve a current controversy, we present here an in-depth study of charge transport properties of Au|alkanedithiol|Au junctions. Conductance histograms extracted from our STM measurements unambiguously confirm features showing more than one set of junction configurations. Based on quantum chemistry calculations, we propose that certain combinations of different sulfur-gold couplings and *trans/gauche* conformations act as the driving agents. The present study may have implications for experimental methodology: whenever conductances of different junction conformations are not statistically independent, the conductance histogram technique can exhibit a single series only, even though a much larger abundance of microscopic realizations exists.

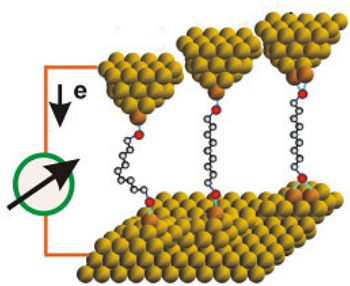

---

[*]Corresponding authors: Thomas.Wandlowski@dcb.unibe.ch; Ferdinand.Evers@int.fzk.de.

# Introduction

The ability to measure and control electrical currents through single molecules offers unique opportunities for exploring charge transport processes in tailored nanoscale junctions. Current approaches fall into three categories: scanning probe methods [1,2,3,4,5,6,7], supported nanoelectrode assemblies [8,9,10,11] and mechanically controlled junctions, which have been refined, so that gaps with even sub-Angstrom resolution can be created using a piezo electric transducer or other mechanical actuation mechanisms. Specific techniques include mechanically controlled break junctions (MCBJ, [12, 13, 14, 15]), STM and conducting AFM break junctions [16, 17] as well as STM-controlled point contact measurements [18]. The latter group of methods is particularly attractive because the technique is applicable to different environments, such as gas and solution phases as well as UHV conditions. Approaches differ in the following criteria: (i) formation of reproducible contacts between a molecule and two probing electrodes, (ii) access to "signatures" of single molecules, (iii) details of statistical data analysis.

In order to improve experimental reproducibility, Xu and Tao [16] have begun to analyze the statistical properties of conductance traces, which show the conductance over a control parameter, measuring the electrode separation. Such current-distance traces were obtained from a repeated formation and breaking of a large number of individual molecular junctions formed between a gold STM tip and a gold substrate in a solution containing molecules [16]. The procedure was adopted and modified by several other groups [18-26]. The statistical interpretation of the individual conductance-distance traces is mostly based on the analysis of conductance plateaus [16, 18, 19, 23, 25] that leads to the construction of conductance histograms. Alternative approaches involve the logarithmic analysis of entire transients [21, 27, 28] or the rapid conductance drop in the last step [22].

Despite using the statistical analysis and comparable experimental techniques, Tao *et al.* [16] and Haiss *et al.* [29] obtained qualitatively different traces for the average conductance $G(n)$ of an *n*-alkanedithiol: while Ref. 16 reports an exponential dependence $G(n) \sim \exp(-\beta_N n)$ with $\beta_N \approx 1$, data in Ref. 29 were represented by an exponential law with $\beta_N = 0.52$, and are smaller by more than an order of magnitude. Slowinski *et al.* [25] also confirmed the exponential behavior. They found $\beta_N \approx 0.83$. However, after refinement of their measurement technique, Tao, Lindsay and coworkers report [30], that in fact the conductance histogram analysis would not yield a unique trace $G(n)$ but rather two traces and none of them coinciding with the experiment by Haiss *et al.* [29]. According to Fujihira *et al.* [23], this result is still incomplete, even a third trace should exist.

The experimental controversy may seem surprising at first sight, since actually the stretched (*trans* configuration) alkane wire is a reasonably simple system: it has an even number of electrons per unit cell and thus is a band insulator. Since the alkane work function is only ~2eV below the gold Fermi energy ($E_F$), while the alkane's conduction band is more than 5eV above (according to our calculations, see below), the coupled wire shows HOMO mediated transport. The tunneling barrier height $\Phi_B$ is given by the energy difference of $E_F$ and the wire's upper valence band edge: $\Phi_B = E_F - E^*_{HOMO}$.

The variability of experimental results reflects fluctuations in the microscopic junction conditions. Therefore, variations in the metal-electrode contact geometry [21, 30, 31] and isomeric structures [29, 32] (*gauche* conformations) have been put forward as an explanation for the appearance of multiple experimental conductance traces. However, it is clear that such an explanation is rather incomplete. Experimentally, a great many combinations of contact/isomeric degrees of freedom could exist, and in the case of the alkanes a substantial fraction of them should contribute to the transport characteristics. Since such junction modifications might be expected to exhibit conductances somewhat different from each other, it is in fact a nontrivial observation and an intriguing open question why equally spaced peak structures in the conductance histograms can appear at all.

This paper constitutes an attempt to resolve this puzzle and to present a unifying view. We report an experimental STM study, which establishes the existence of at least three series of peaks in conductance histograms of α,ω-alkanedithiols. Based on *ab initio* simulations of the single molecule conductance we will argue, that many junction configurations add to the conductance histograms, as expected. Furthermore, our calculations also show that conductances of different junction conformations are correlated and in fact cluster at roughly integer multiples of some basic units. For this reason the equally spaced peak structures in conductance histograms are not smeared out. This illustrates an important conceptual consequence: in contrast to a wide spread practice, a single peak sequence cannot necessarily be attributed to a specific junction configuration but rather it also can – and in the case of the alkanes it does – result from averaging over many junction configurations.

## Experimental

### *Chemicals*

The α,ω-alkanedithiols 1,5-pentanedithiol (PD), 1,6-hexanedithiol (HD), 1,8-octansdithiole (OD), 1,9-nonanedithiol (ND) and 1,10-decanedithiol (DD) were purchased from Aldrich (reagent grade) and used without further purification. The electrolyte solutions were prepared with Milli-Q water (18 MΩ, 2ppb TOC) and HCl (35%, Merck suprapure). Absolute ethanol (KMF 08-205) and 1,3,5-trimethylbenzene (TMB, p.a. 98%) were obtained from KMF Laborchemie Handels GmbH and Sigma-Aldrich, respectively.

### *Electrode and sample preparation*

The Au(111) electrodes used in this work were discs of 2 mm height and 10 mm in diameter, nominal miscut < 0.1°. Island-free Au(111)-(1x1) surfaces were prepared by immersing a flame-annealed Au(111)-(p x √3) electrode into deareated 0.05 M HCl for 20 min [33]. The electrode was then emmersed, rinsed with absolute ethanol and dried in a stream of argon.

The organic adlayers were prepared by immersing of the dry Au(111)-(1x1) electrode into a 0.1 mM ethanolic solution of the respective α,ω-alkanedithiol for 2 min. Subsequently, the modified electrodes were removed from the assembly solution, rinsed with warm ethanol, transferred into the STM cell and covered with TMB. Selected experiments were carried out directly in 0.1 mM TMB solutions of the α,ω-alkanedithiols.

### *Single Junction Conductance Measurements*

The STM imaging and the STS (current-distance) measurements were carried out sequentially at room temperature in 1,3,5-trimethylbenzene (TMB) using a modified Molecular Imaging PicoSPM. The STM tips were uncoated, electrochemically etched gold wires (99.999 %, 0.25 mm diameter, etching solution 1:1 mixture of 30 % HCl and ethanol [34]. All STM measurements were conducted in a sealed container, to prevent oxygen exposure, and in constant current mode employing low tunneling currents (10 pA to 100 pA), and bias voltages $E_{bias}$ ranging between ± 0.010V and ± 0.300V.

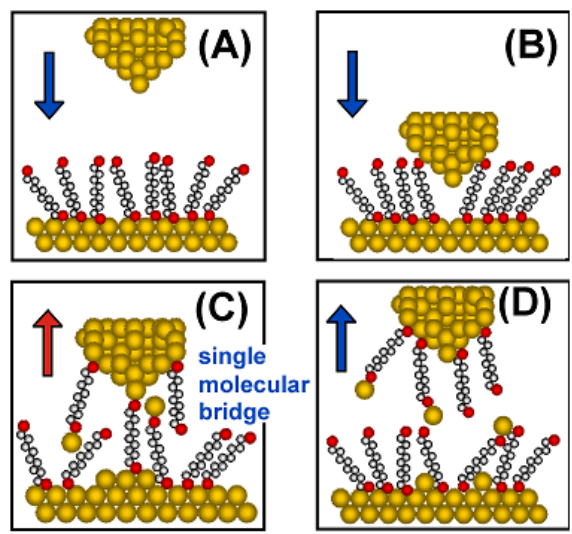

**Figure 1:** Schematic representation of the STM-type "contact junction" approach applied in the present study. (A) approach, (B) formation of molecular contacts, (C) pulling and (D) breaking.

Single junction conductance data were obtained from current-distance traces employing either a single channel preamplifier (0.1 nA/V, 1.0 nA/V or 10 nA/V) or a dual multi-channel preamplifier stage capable to record currents in 1 pA $\leq i_T \leq$ 210 µA simultaneously [35]. The latter was combined with a customized control circuit and a multi-channel scope recorder Yokogawa DL 750 (1 MSs$^{-1}$, 16 bit). The following sequence was applied (Fig. 1): A sharp gold STM tip, capable for imaging experiments with atomic resolution, was brought to a preset tunneling position, typically defined by $i_T$ = 50 or 100 pA and $E_{bias}$ ranging between ± 0.010 V and ± 0.300 V (Fig. 1A). Subsequently, the STM feedback was switched off, and the tip approaches the adsorbate-modified substrate surface at constant x-y position (Fig. 1B). The approach was stopped before reaching point contact with the gold surface ("gently touch") by choosing an upper threshold current of 0.2 $G_0$ $E_{bias}$ ($G_0$ = 77 µS). These setting still ensure rather strong interactions between the gold tip and the α,ω-alkanedithiol adlayer. After a dwelling time of 100 ms, sufficient to create molecular junctions between tip and substrate, the tip was retracted with a rate of 5 to 10 nm s$^{-1}$ until a lower threshold position was reached. The latter was chosen sufficiently far from the adsorbate-modified substrate ensuring the breaking of all Au│α,ω-alkanedithiol│Au junctions. This process may also involve the attachment/ detachment and/or reorganization of gold adatoms [17, 19, 36, 37] The transient conductance curves were recorded, and the entire cycle was repeated up to 3000 times.

We notice that the experimental technique applied in the present study exhibits a distinct difference compared to the original approach reported by Xu *et al.* in Ref. 16. The choice of a single or dual preamplifier stage with a wide dynamic range and coupled with a high-sensitive feedback loop [35] ensured that no physical contact between gold tip and gold substrate occurred prior to the formation of molecular junctions. This new technical development increased stability and reproducibility of the single molecular junction studies with alkanedithiols significantly. The "gentle touch" technique does not lead to major modifications of the STM tip due to tip crashes. In consequence, the adsorbate-covered surface could always be monitored between cycles of current-distance traces with a high quality tip capable of molecular resolution.

## Results

Fig. 2A illustrates, as an example, the typical response of the simultaneously recorded high and low current channels during a complete approaching/retracting cycle for 0.1 mM 1,9-nonanedithiol (ND) in TMB. The upper current limit of 1.5 µA, as monitored in the high current channel, prevents the formation of Au-Au nanocontacts. STM images, recorded with the same gold tip before and after the transient conductance trace, are identical exhibiting either an ordered low-coverage striped phase (Fig. 2B) or a disordered high coverage phase with characteristic vacancy islands (Fig. 2C) depending on the initial assembly conditions. Occasionally, a few monatomic gold islands were observed after a long series of current distance traces (c.f. inset in Fig. 2C).

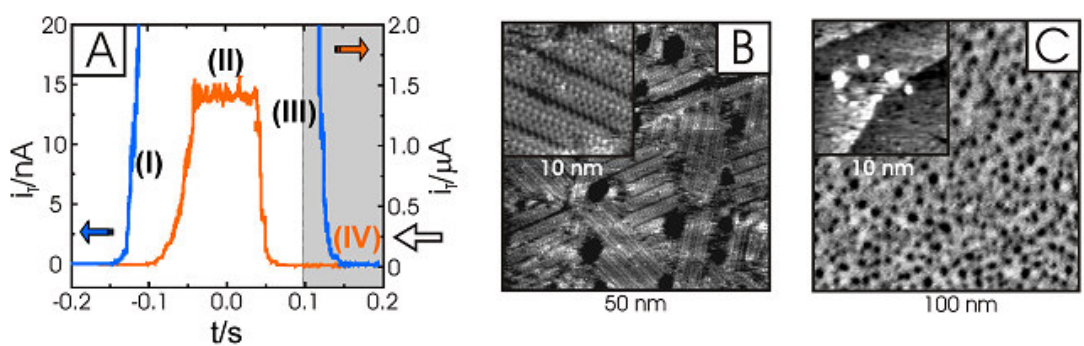

**Figure 2:** (A) Simultaneously recorded current *vs.* time traces of a complete approaching and breaking cycle for 0.1 mM 1,9-nonanedithiol (ND) in 1,3,5-trimethylbenzene (TMB). The symbols of the individual steps follow the nomenclature of Fig. 1. The orange trace represents the high current channel (maximum range 3 nA < $i_T$ ≤ 23 µA), which monitors the exact position of the tip relative to the substrate surface. Typically, we stopped the z-movement of the tip upon reaching a current between 1 and 2 µA. The orange trace shown in Fig.2A refers to $i_T$ ~ 1.5 µA. The blue trace illustrates the low current channel (10 pA < $i_T$ < 35 nA), which is chosen to follow single junction conductance characteristics. (B) Large scale and high resolution (inset) STM images of the striped low coverage phase ($i_T$ = 0.1 nA; $E_{bias}$ = 0.1 V ) of ND on Au(111)-(1x1) in TMB monitored with a gold STM tip before and during a sequence of stretching experiment. (C) Large scale image of the disordered high coverage phase of ND on Au(111)-(1x1), other conditions identical to (B). The adlayers are typically unaltered. Only occasionally we observed the formation of small, monatomically high gold islands.

The retraction or pulling curves were recorded with high resolution in 1 pA < $i_T$ < 100 nA. We observed three types of transient curves. Type I curves are exponential and represent direct electron tunneling between the gold tip and the substrate without molecular junctions being formed [26, 30]. The percentage of these decay curves was 50% when performing the experiment in TMB containing 0.1 mM ND, and up to 80% in the presence of an ND adlayer without molecules being dissolved in electrolyte. Assuming a rectangular barrier, we estimated at large tip-adsorbate distances an effective barrier height $\Phi_{eff} = (1.1 \pm 0.2)$ eV, which is comparable with data obtained for the bare gold│TMB system [38]. Type II curves are non-monotonic and noisy, which could be attributed to mechanical instabilities. The percentage of these curves was around 10 %. Traces of type I and II were rejected in the further analysis of the experimental data. The remaining type III conductance traces are monotonic and non-exponential. Fig. 3 illustrates typical examples for 0.1 mM ND in TMB recorded with preamplifier settings of 0.1 nA V$^{-1}$ (A) and 1.0 nA V$^{-1}$ (B) for the low current channel. They exhibit single plateaus or series of plateaus with a typical length of 0.04 to 0.15 nm, which are separated by abrupt steps of 50 pA up to 3 nA. These current steps are assigned to the breaking of individual respective multi-molecular ND junctions previously formed between the gold STM tip and the substrate surface [26, 30]. The percentage of type III conductance traces varied between 40 % and 10 % depending if ND is dissolved in TMB or only pre-adsorbed on the Au(111)-(1x1) surface. Similar observations were reported previously by Li *et al.* [30]. Control experiments in pure TMB displayed almost exclusively (98 %) exponentially decaying traces. No evidence of current steps resembling type III traces was found.

The statistical analysis of type III conductance traces was carried out by constructing histograms, typically based on up to 1500 individual curves. We have chosen a numerical routine, which selects plateau currents according to the following criteria [19, 38]: minimum plateau length of 0.04 nm and an average variation of the current magnitude of less than 5%. Fig. 4 displays two histograms, which were obtained from two experiments with 0.1 mM ND in TMB under conditions illustrated in Fig. 3. An analysis of the characteristic conductance peaks allows identifying three different sequences of equally spaced maxima. They are attributed to low, L, medium, M, and high, H, conductance junctions. The current within each series scales *approximately* linearly with the number of peaks. The first peak of each sequence is attributed to a single molecular junction. The corresponding currents depend also linearly on the applied bias voltage, at least up to ± 0.200 V (c.f. insets in Fig. 4A and 4B). These correlations lead to the following conductance values of the three specific single molecule Au│ND│Au junctions: (0.47 ± 0.03) nS (L), (2.0 ± 0.2) nS (M) and (9.9 ± 0.9) nS (H). We obtained the same results when constructing all-data-point histograms (c.f. Fig. S1 in Supporting Information). A comprehensive discussion of advantages and pitfalls of different approaches for the construction of conductance histograms will be presented in a forthcoming paper [38].

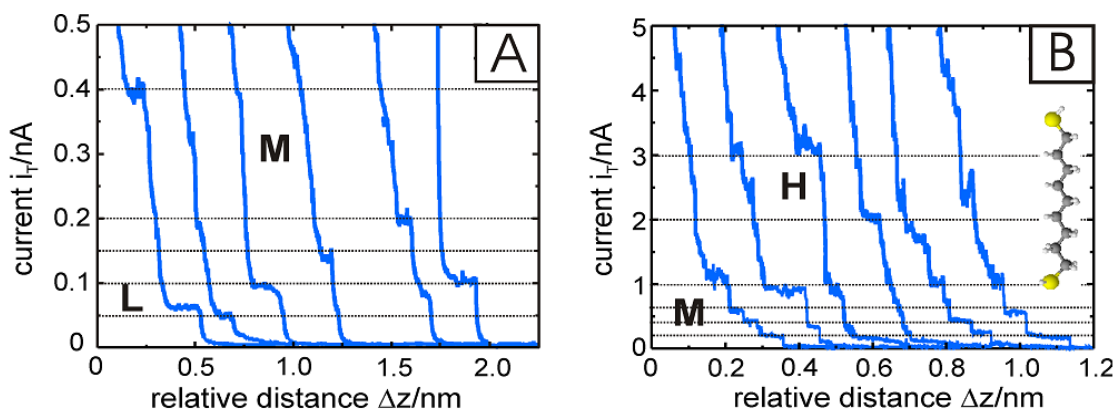

***Figure 3:*** (A) Current – distance retraction curves recorded with a gold STM tip (low current channel with a preamplifier limit of 1 nA) for 0.1 mM 1,9-nonanedithiol in 1,3,5-trimethylbenzene on Au(111)-(1x1), at E$_{bias}$ = 0.10 V. The setpoint current before disabling the feedback was chosen at i$_o$ = 100 pA. The pulling rate was 4 nm s$^{-1}$. (B) Same conditions as in (A), except that the preamplifier limit was chosen at 10 nA. The dotted lines represent characteristic regions of the low, mid and high currents. For further explanation we refer to the text.

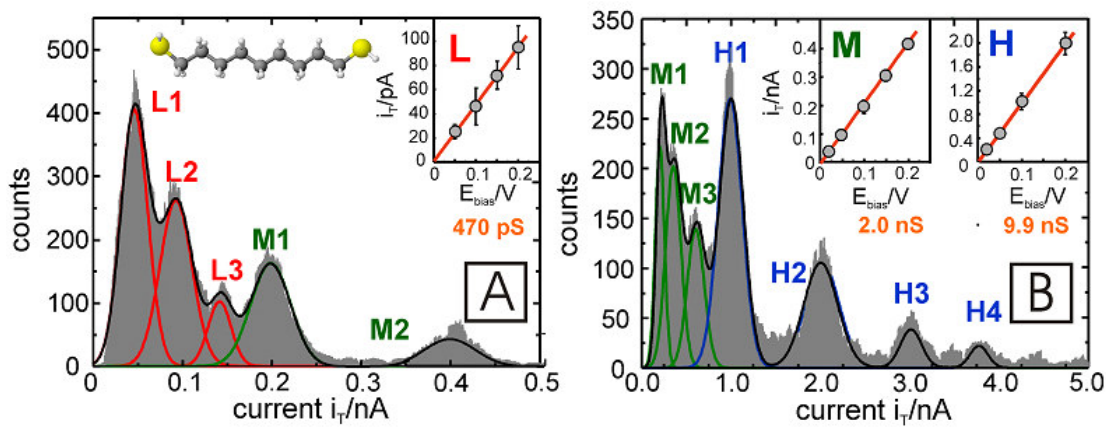

***Figure 4:*** Conductance histogram constructed from values of the plateaus of type III "stretching" curves for Au│1,9-nonanedithiol│Au junctions. (A) 1600 out of 4300 traces employing the 1 nA (max) preamplifier; (B) 1100 out of 4300 traces recorded with the 10 nA (max) preamplifier. All other conditions are identical to those in Fig. 3. We notice that the low conductance sequence could only be resolved with the high sensitive preamplifier. The insets in (A) and (B) show that the current within each series scales *approximately* linearly with the number of peaks.

Following the same experimental protocol we obtained also multi-peak conductance histograms for 1,5-pentanedithiol (PD), 1,6-hexanedithiol (HD), 1,8-octanedithiol (OD) and 1,10-decanedithiol (DD) in TMB. The respective data of the single junction conductances are summarized in Tab. 1.

The values of the H - conductance are approximately a factor of 5 larger compared to the M - conductance for the same molecule, while the lowest values L do not scale with a constant ratio. The M - and L – conductance data approach each other with increasing chain length. As an example, we refer to M(PD) : L(PD) ~ 34 and M(DD) : L(DD) ~ 2. Furthermore, we found that the single junction conductance data of the three sequences are rather independent on alkanedithiol concentration and coverage. No preferential occurrence of H, M or L conductance junctions was observed for the aliphatic molecules studied. However, within one sequence one could typically resolve up to three peaks with a distinct preference for single molecular junctions. Integration yields an average ratio of 0.54 : 0.35 : 0.11 to form one-, two- or three-molecule junctions.

***Table 1:*** Single junction conductance data Au│α,ω-alkanedithiol │Au.

| α,ω-alkanedithiol | L – conductance (nS) | M – conductance (nS) | H – conductance (nS) |
|---|---|---|---|
| PD (n = 5) | 1.9 ± 0.05 | 64 ± 5 | — |
| HD (n = 6) | 2.45 ± 0.06 | 20 ± 2 | 95 ± 10 |
| OD (n = 8) | 0.89 ± 0.08 | 4.4 ± 0.4 | 21 ± 2 |
| ND (n = 9) | 0.47 ± 0.03 | 2.0 ± 0.2 | 9.9 ± 0.9 |
| DD (n = 10) | 0.22 ± 0.02 | 0.45 ± 0.04 | 1.68 ± 0.03 |

We also note that the H- and M – values of the single junction conductance for HD, OD and DD, as obtained in the present study, are in agreement with data recorded after breaking preformed Au-Au nanocontacts in toluene containing 1.0 mM solutions of the respective alkanedithiols [16, 22, 30]. The mid-conductance of a single molecular Au│OD│Au junction is supported by experiments of Gonzalez *et al.* [28], who carried out a MCBJ experiment in a liquid cell filled with 1 mM OD in TMB. The values of the L – conductance of HD, OD and ND are comparable with data reported by Haiss *et al.* in air, toluene as well as under electrochemical conditions [29, 32]. No agreement was found between the results of our study and the STM – based single molecule conductance measurements of Fujihira *et al.* carried out under UHV conditions for Au│HD│Au [23].

Fig. 5 shows the semi-logarithmic plot of the conductance versus molecular length. The latter is expressed as the number of $CH_2$ - units. The H- and M conductance values follow a simple tunneling model given by $G = G_c \cdot \exp(-\beta_N \cdot n)$ with decay constants $\beta_N$ of $(0.96 \pm 0.15)$ and $(0.94 \pm 0.05)$, respectively. These values of $\beta_N$ are in agreement with literature data on the M and/or H conductances of single junctions Au│alkandithiol│Au [16, 22, 30] as well as for the electron transfer through compact and aligned monolayers of alkanethiols using nanopores [39], mercury contacts [40, 41, 42], CP-AFM [43] or redox probes [44, 45].

The low conductance data give a rather poor linear correlation with a decay constant $\beta_N \sim (0.45 \pm 0.09)$, distinctly different from the H- and M- sequences. However, we notice that the estimated value $\beta_N$ is rather close to that reported by Haiss *et al.* [29].

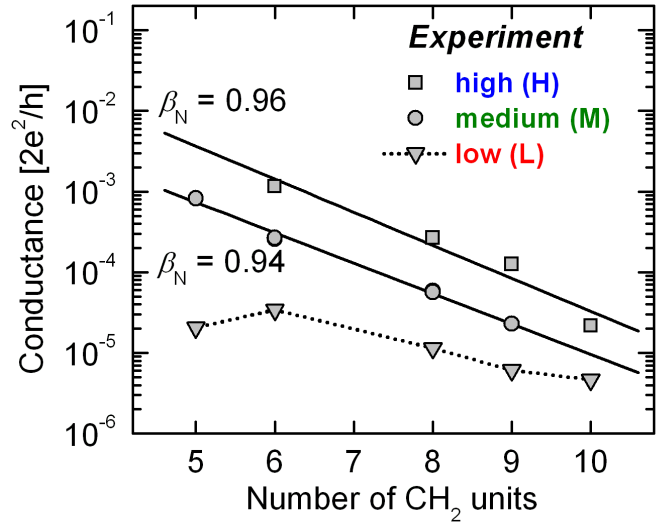

***Figure 5:*** Chain length dependence of the single junction Au-alkanedithiol-Au conductivity in a semilogarithmic representation. The three sets of conductance values – high (H), medium (M) and low (L) – are shown as squares, circles and triangles. The straight lines were obtained from a linear regression analysis with decay constants $\beta_N$ defined per methylene ($CH_2$) unit.

# Theoretical

## *Model and Method*

To provide insight into the multiple peak series observed experimentally, we have performed DFT based calculations of alkanedithiols coupled to Au electrodes, using a standard quantum chemistry package TURBOMOLE [46]. Calculations were done for different configurations of an "*extended molecule*" composed of an *n*-alkanedithiol of variable chain lengths ($n = 4 \ldots 10$) bridged between two pyramids of ~ $40 \div 55$ Au atoms (Fig.6). These clusters model the contact region of the gold electrodes. We employed the generalized gradient approximation (GGA, BP86 functional [47]) with contracted Gaussian-type basis set of triple-zeta valence quality including polarization functions [48]. The experimental lattice constant of 4.08 Å was fixed for bulk fcc-Au clusters, whereas relaxed configurations were found for the alkanedithiol molecules coupled to few gold atoms assuming different gold - sulfur coordination geometries, namely: (i) S atop a Au atom (Fig. 6A,B); (ii) S forming a bridge with two Au atoms (Fig. 6C) or (iii) S placed in a (111) hollow site [49]. We obtained S-Au bond distances of 2.28 Å, 2.36 Å, and 2.72 Å for atop, bridge and hollow-site configurations, respectively. We also considered all-trans and various gauche conformations of the alkyl chains. Conductances have been calculated within the Landauer approach, as implemented in a homemade simulation package [50, 51, 52].

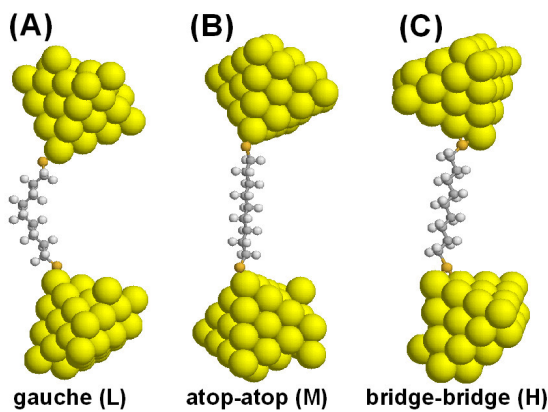

**Figure 6:** Three typical arrangements of a single alkanedithiol molecule bridged between Au electrodes as used for the conductance calculations. (A) 1,9-nonanedithiol (ND) with one gauche defect and both terminal sulfur atoms coordinated in atop position (low, L), (B) ND in all-trans conformation and atop-atop coordination (medium, M), (C) all-trans ND in a bridge – bridge coordination (high, H).

## *Results*

**Elementary consideration of transport mechanism**. To begin with, we consider all-trans isomers with the sulfur coupled to a single Au atom at each electrode ("atop-atop" geometry, Fig. 6B). For the description of the electron transport, we consider the energy-dependent transmission $T(E)$ (Fig.7A). It represents the probability for electrons injected with the energy $E$ from one electrode to be transmitted through the molecular junction. The conductance is defined by the transmission $T(E_F)$ evaluated at the Fermi energy ($E_F$) in units of the conductance quantum $G_0 = 2e^2/h = 77.5$ μS. Molecular states of alkanedithiols appear as resonance peaks in the transmission spectrum (see Fig.7A).

Our findings fully support the qualitative picture of transport through alkanes outlined in the introduction. The Fermi energy $E_F$ is situated inside the alkane HOMO-LUMO gap (7.5eV in the present calculations) giving rise to an effective barrier $\Phi_B = E_F - E^*_{HOMO} = 2.14$ eV. The current flow involves mainly the alkane-HOMO (HOMO* of the alkanedithiol) depicted in Fig.7A (insert). Accordingly, the transmission $T(E)$ drops down rapidly around -2.2 eV below $E_F$ and a nearly insulating gap spreads up all the way until 5.3 eV above $E_F$ (alkane-LUMO, Fig. 7A). Furthermore, the conductance through $n$-alkanedithiol junctions decays exponentially according to $G(n) = G_c \exp(-\beta_N n)$ with a decay constant $\beta_N \propto \Phi_B^{1/2} d_0$, where $d_0 = 1.28$Å is the unit length of the alkane chain. We found $\beta_N = 0.83$ per $CH_2$ unit (open circles in Fig. 8A). A value $G_c = 0.24 G_0 = 18.5\mu S$ was obtained for the contact conductance.

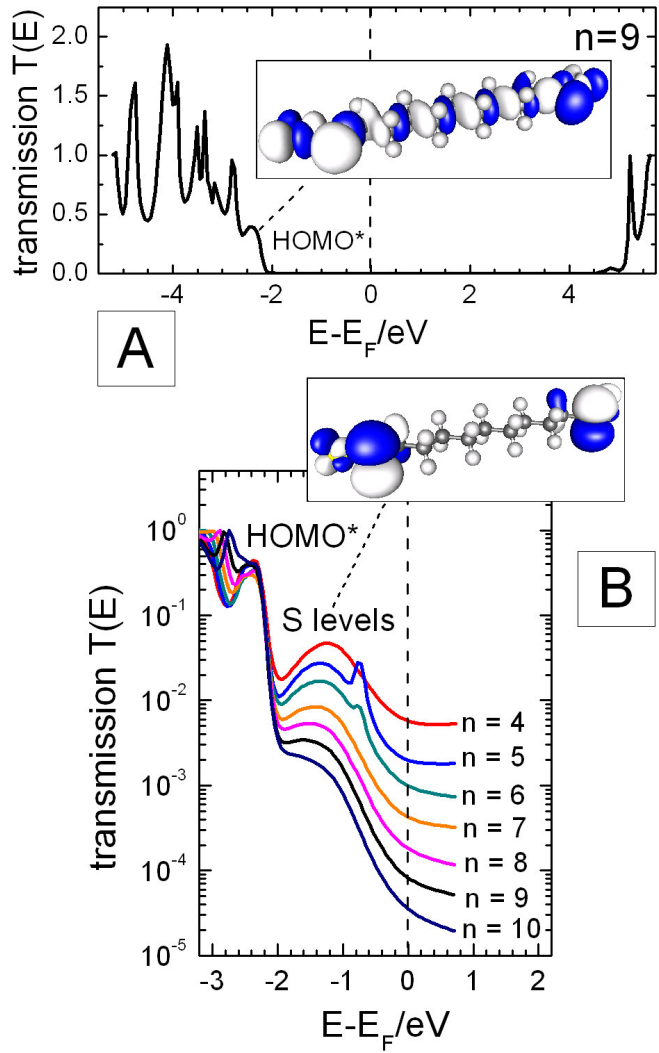

**Figure 7:** (A) Calculated energy-dependent transmission $T(E)$ for a single 1,9-nonanedithiol molecule ($n=9$) bridged between gold electrodes in an all-trans "atop-atop" arrangement [panel (B) in Fig. 6]. The inset plot shows a molecular orbital of the Au-S-$(CH_2)_9$-S-Au cluster, formed by the nonane HOMO hybridized with the S and the Au electronic levels. This state determines the tunneling barrier. (B) Energy-dependent transmissions of the $n$-alkanedithiols of various lengths $n$ around the Fermi level. The inset displays the alkanedithiol's HOMO, which is an antisymmetric superposition of atomic wave functions localized on both sulfur atoms. The energy difference within its symmetric part, HOMO-1, is exponentially decaying with $n$.

So far, our analysis has not yet included the thiol end-groups. They introduce two evanescent gap states built of symmetric and antisymmetric combinations of wave functions localized at the *sulfur* atoms (Fig. 7B, inset). These states appear as broad resonances centered around $E_S \approx -1.4$eV below $E_F$ in the transmission spectrum (see Fig. 7B). To describe such a situation, we apply the Breit-Wigner formula [53], $T(E) = 2\Gamma_S \gamma_n(E)/[(E-E_S)^2+(\Gamma_S+\gamma_n(E))^2/4]$. Here, $\Gamma_S$ denotes an on-resonance probability amplitude (inverse life-time) for an electron to hop from the S atom to the nearest electrode. From our data we obtain $\Gamma_S = 1.25$eV. The other probability $\gamma_n(E) \ll \Gamma_S$ reflects tunneling through the molecule and is exponentially small, $\gamma_n(E)=\gamma_0\exp[-\beta(E)n]$, where $\beta(E) \propto [E-E^*_{HOMO}]^{1/2}d_0$. The tails of such broad peaks, when approaching the Fermi energy (Fig. 7B), define the amplitude of the molecular conductance $G(n) \cong 2G_0\Gamma_S\gamma_0\exp[-\beta(E_F)n]/(E_F-E_S)^2$. Notice, that an evanescent state localized close to an Au surface does not change the tunneling asymptotic, which is defined by an offset of the alkane HOMO state ($E^*_{HOMO}$) with respect to $E_F$, since $\beta(E_F) = \beta_N$.

**Effect of contact geometry.** The effect of contact geometry was appreciated by many authors [31, 50, 54, 55, 56]. Fig. 8A clearly shows that many contact configurations exhibit comparable conductance values. Largest values (squares, Fig.8A) are associated with all-trans configurations and the sulfur bound to two Au atoms on either side (S-Au$_2$, "bridge-bridge" configuration, Fig. 6C). For a double S-Au$_1$ contact configuration ("atop-atop", Fig.6B) the conductance is reduced by a factor of 4, i.e. a factor of 2 per each anchoring group (circles, Fig.8A). The combination S-Au$_2$ on one side and S-Au$_1$ ("bridge-atop") on the other predicts a factor of two. A rational for the factors of 2 per anchoring group is readily found by recalling that the number of escape channels per side doubles ($\Gamma_S \rightarrow 2\Gamma_S$), when the number of Au atoms connecting to the sulfur increases from 1 to 2 [57]. In addition, we explored configurations with alkanedithiols embedded in short chains of gold adatoms ("extended atop" geometry). The resulting values of the junction conductances were changing within 10% only, and therefore such configurations do not need to be considered separately. Contradicting statements in the literature [58] according to which adatoms would change conductances of S-Au bonding molecules by an order of magnitude suffer heavily from modeling artifacts. Specifically, these earlier calculations assumed an unphysical angle of 180 degrees of the Au-S-C angle, while the correct value used in our analysis is $\approx 104.5$ degrees.

**Effect of gauche defects.** We have repeated similar calculations for different gauche-isomers. Because of the structural defect in the otherwise homogeneous potential barrier, they exhibit conductance values typically a factor ~10 smaller than those of all-trans alkanedithiol chains (see Fig. 8A, triangles). Also the different gauche-isomers can be coupled to Au leads via different contact geometries. The net effect of the combined degrees of freedom, gauche and contact geometry, is that the molecular junctions with alkanedithiols can exhibit conductance values with a spreading of roughly two orders of magnitude.

**Conductance histograms.** Referring to the above situation, it is not clear *a priori*, which result the experimentally used conductance histogram technique would actually show. We mimic the experimental approach and simulated conductance histograms to obtain a rough idea (Fig. 8B). Essentially, we assume that all single molecule contact configurations occur with equal probability and molecular bridges up to four molecules (*trans* or *gauche* conformations) can be formed. However, the probability of *n*-molecular junction to occur is suppressed by a factor of $p^{n-1}$ where $p < 1$. In other words, we assume that the probability for a contact with two molecules in parallel is a factor of $p$ lower than for a single molecule contact, etc. We propose that thermodynamic fluctuations imply an incoherent averaging of conductance values associated with conformational and other environmental degrees of freedom. The associated peak broadening we model by a Gaussian distribution assigned to each calculated conductance value shown in Fig. 8A (see Ref. 59 and Supporting Information for further details). We believe that as rough as our statistical model may be, its qualitative features are significant since they are not too much dependent on specific model assumptions.

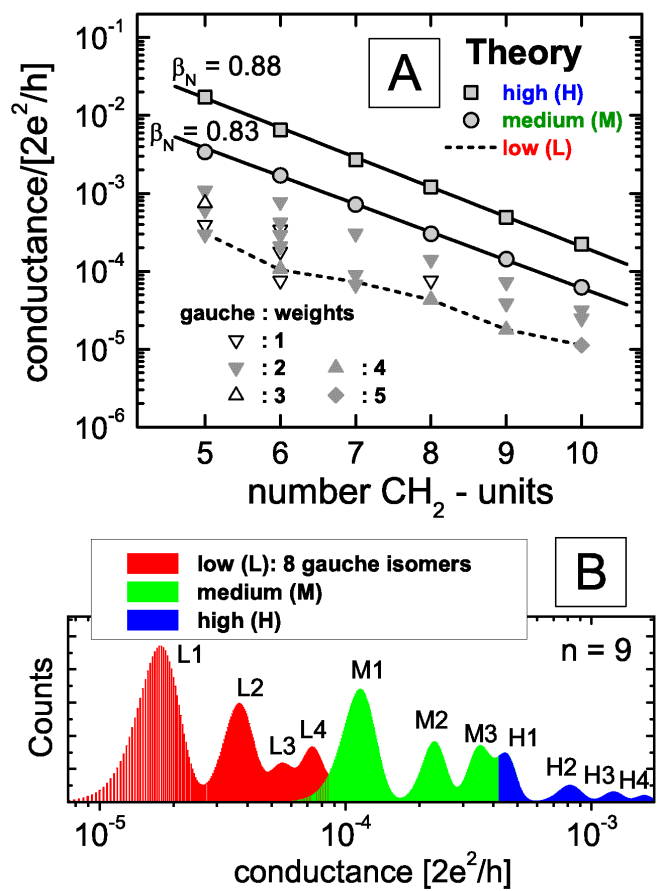

*Figure 8:* (A) Calculated conductance values of the Au|*n*-alkanedithiol|Au junctions *vs* number *n* of methylene units in a semilogarithmic representation. The straight lines are based on the linear regression analysis. The high (H, filled squares) and medium (M, filled circles) values were directly obtained for the arrangements of molecular bridges depicted in Fig. 6C and Fig. 6B. The conductances of many different, nonequivalent *gauche* isomers cover the window below the medium values. An alkyl chain obeys mirror symmetry. Therefore, depending on the position of the structural defect (c.f. Fig. 6A) different configurations can have the same conductances. Thus, statistical weights, 1 or 2, can be ascribed to calculated values. Since some of the obtained values were very close to each other, weights up to 5 appeared as well. (B) Example of a simulated conductance histogram for a molecule with *n* = 9 units. Extracting the positions of the lowest conductance peaks leads to the dashed line drawn in Fig. 8A. Further details of the histogram construction are given in Ref. 59 (Supporting Information).

The key result of this analysis is that three series of conductance peaks, L, M and H, may be identified from the simulated conductance histograms (Fig. 8B). The peak values are nearly equally spaced, neatly separated by integer multiples of some "master values". The L-series originates from gauche-isomers. The M-series is due to all-trans molecules coupled to single Au atoms at one or both sides. The H-series comes from the all-trans isomers where both S atoms are coordinated with two Au atoms. Notice, that the L-peaks do not strictly follow an exponential *n*-dependence, which is not surprising since they represent a convolution of many different junction conformations. By contrast, calculated "middle" (M) and "high" (H) conductance values fit to the exponential decay law with $\beta_N$ = 0.83 to 0.88 per CH$_2$ unit.

**Decay constant $\beta_N$.** The calculations of $\beta_N$ were carried out using the generalized gradient approximation, GGA. In order to estimate the error related to the local character of GGA, we have performed additional calculations with a hybrid functional, B3LYP, which partly captures the long-range component of the exact functional that is completely absent in local approximations. The calculation within the B3LYP functional [60] leads to an increase of the HOMO-LUMO gap for the pure alkane chains and to a shift of the alkane-HOMO down to -3.1eV with respect to $E_F$ that implies a higher tunneling barrier $\Phi_B$. Since $\beta_N \propto \Phi_B^{1/2}$, the decay constant $\beta_N$ increases from 0.83 to 1.0, as it should. Consistent with this observation, another hybrid functional, TPSSH [61], also leads to an increased $\beta_N \approx 0.93$. Based on this error analysis, we believe that the exact value could be up to ~20% larger than our GGA based result.

The estimates based on a complex band structure analysis, performed by Tomfohr and Sankey [62], and by Picaud *et al*. [63], suggested $\beta_N$ ~1.0 and ~0.9 respectively, in agreement with our findings. Their estimates for the tunneling barrier $\Phi_B$ ~ 3.5 ÷5.0 eV, however, exceed our value 2.1 by a factor of ~2. Another study by Müller [31] reported a comprehensive transport calculation using the TRANSIESTA package. The qualitative findings on the dependence of the conductance on contact geometry are consistent with our observations. However, a serious discrepancy prevails in the exponent, which was $\beta_N$ = 1.25 in Ref. 31. In that study, the exponent together with the HOMO-LUMO gap, ~17eV, exceed considerably well-established experimental values ($\beta_N$ ~ 1.0 and gap ~8.5 ÷ 10.0 eV [64]), which is hard to reconcile within a general expectation of LDA/GGA. An interesting attempt to go beyond the DFT approach for the conductance through alkanes was made by Fagas *et al*. [65] using a configuration interaction method. Unfortunately, the discrepancy between the obtained value, $\beta_N$ = 0.5, and the experimental observation is even larger, than the DFT related uncertainty.

## Discussion

Comparing Figs. 5 and Fig. 8A, we conclude that the *qualitative* agreement between experiment and theory is very good. Theory confirms the existence of three peak series in the experimental conductance histograms. The comparison of experimental observations and quantum chemical calculations suggests the following assignment: The M-series of junction conductance data is attributed to an all-trans conformation of the alkyl chain and a double S-Au$_1$ contact geometry. The H-series is assigned to all-trans alkyl chain with the sulfur bound to two Au atoms on either side (S-Au$_2$ bridge-bridge geometry). The junction conductance between the two configurations differs by a factor of four. Experimentally a factor of ~5 was observed.

We expect, that also mixed couplings (S-Au$_2$ on one side and S-Au$_1$ on the other) yielding a factor of roughly two in the conductance occur. An associated series is not necessarily easily distinguished experimentally, since its conductance peaks coincide with peak positions of the M and H series. We interpret the absence of clear indications for a mixed series in the experimental data as an indication, that the occurrence of mixed contact geometries is somewhat discouraged against the others by the experimental boundary conditions. Since only the all-trans conformation of the alkyl chains is involved for the M- and H series of the junction conductance data, the decay of the conductance with the chain length is exponential with a theoretical decay constant $\beta_N$ = 0.83 to 0.88 per CH$_2$ unit, which is very close to our experimental one $\beta_N$ = 0.94 to 0.96. The latter value ~1 has been confirmed by a large number of experiments by now with single alkanedithiols as well as with alkanethiol monolayers [16, 22, 30, 39-45].

By contrast, the L series is much more complicated (Fig. 4, Ref. 29, 32). Our theoretical analysis suggests that gauge sites represent structural defects, which can lower the junction conductance up to a factor of 10 compared to all-trans alkyl chains. This trend is qualitatively not modified when assuming different contact geometries including short chains of gold adatoms. The experimentally observed ratio between M- and L-conductance data ranges between 34 for PD down to ~ 2 for DD. In consequence, we assign these L-conductance data to molecular junctions of various contact geometries with alkyl chains having at least one gauche defect. This manifests itself in a behavior of $G(n)$ which does not follow a simple exponential law. Additional support of our interpretation of the L-series as an average over a certain ensemble of contact and gauge geometries might come from temperature dependent transport measurements, Ref. 32. For the H and M series one may expect, that the leading $T$-effect related to geometry is a modification of the relative probability to find an H vs. M series. In fact, at sufficiently high $T$, an experimental current-distance curve (with relaxed molecules feeling no external forces) could show an average over different contact geometries, so that it no longer supports a clear plateau structure. By contrast, the L-series should exhibit an additional average over gauche configurations at sufficiently large $T$. While in principle a great many configurations will be sampled, our calculations show that only very few - namely those with the largest conductance - dominate the transport average. As a consequence a roughly exponential dependence of the conductivity, $\sim\exp(-T_0/T)$, might be expected in a certain temperature range. Here, $T_0$ measures the typical energy cost for healing the wire from gauche defect(s). Indeed, this is the exponential dependency that has already been reported before in Ref. 32.

In addition, also the *quantitative* agreement for relative conductance values obtained from theory and experiment is reasonable, which is remarkable since simulations cover a broad range of conductance values. Theory appears to slightly underestimate the values of $\beta_N$, due to the reasons discussed above.

The discrepancy between theory and experiment in absolute values (contact conductance $G_c$) is a factor ~5. Such deviations have been observed frequently before, and have been discussed in Ref. 50. Again, approximations in GGA most likely contribute to the error in the prefactor $G_c$ [66].

Finally, we comment on the occurrence of multiple steps in the measured conductance traces, which could be assigned to different junction geometries. We like to point out that we never observed a M step occurring after a L step or a H step occurring after a M step. The experimental current-distance traces only exhibit transitions from high to lower conductance values. Transitions between H and M steps were observed frequently, while transitions from M steps to a single (or multiple) L step were observed very seldom. One trace, as an example, is plotted in Figure 3A. We suggest that the former represent changes in the contact geometry as the junction is pulled apart. Changes in the molecular conformation appear to be rather improbable. Unfortunately, this hypothesis cannot be tested unambiguously for Au|alkanedithiol|Au junctions. Studies with metal electrodes and anchoring groups allowing only for one contact configuration because of geometrical or electronic reasons could help resolving this open question in more detail.

## Summary and Conclusions

We studied experimentally the conductance of single α,ω-alkanedithiols ($n$ = 5 to 10) bound covalently to two gold electrodes in 1,3,5-trimethylbenzence (TMB) employing an STM-type break junction setup. The extracted conductance histograms revealed characteristic peaks corresponding unambiguously to *three sets* of distinct junction configurations, which were labeled L, M and H. No preferential occurrence was observed under our experimental conditions.

Within each sequence one could typically resolve up to three conductance peaks equally spaced at integer multiples of a fundamental conductance value.

The comparison with quantum chemistry *ab initio* simulations demonstrates that the multiple conductance values of Au│alkanedithiol│Au junctions could be attributed to different Au–sulfur coordination geometries and to different conformations of the alkyl chain. In particular, the "medium" conductance values M correspond to an all-trans conformation of the alkyl chain with one of the sulfur atoms coordinated in atop position to a single Au atom. The "high" conductance values H represent all-trans alkyl chain in combination with both sulfurs coordinated to two Au atoms in bridge geometry. The sequence of "low" conductance values L is attributed to nonequivalent isomers of alkanedithiols with gauche defects.

The experimental data and the *ab initio* quantum chemical simulations demonstrate that the M and H values fit to the exponential law $G(n) = G_c \exp(-\beta_N n)$. The experimental decay constants $\beta_N = 0.94 \div 0.96$ were found to be in good agreement with theoretical data. None-exponential behavior was observed and explained for the L conductance junctions.

We emphasize, that it was the combination of single molecule conductance experiments with first principle calculations that enabled us to develop a comprehensive and uniform understanding of microscopic details of charge transport in Au│alkanedithiol│Au junctions. In this way a proposal has been made, how contradictions between various literature reports [16, 22, 23, 25, 29, 30] could be resolved. We believe that results of our work have important consequences for the use of the conductance histogram technique in systems, where several junctions with comparable conductance values can be realized, as is the case for molecules with isomeric structures. In such cases, peak series may occur which do not correspond to just a single species with single, double, triple, etc. junctions in parallel. Their interpretation may be feasible only with the assistance of elaborate *ab initio* simulations.

## Acknowledgements

This work was supported by the HGF Project "Molecular Switches", the DFG, the Volkswagen Foundation, IFMIT and the Research Center Jülich. I.P acknowledges support of the German Academic Exchange Agency for a PhD fellowship. F.E., A.A. and A.B. acknowledge support from the DFG "Center for Functional Nanostructures" situated at Karlsruhe University. The authors also acknowledge discussions with M. Mayor and A. Goerling.

Additional material on the data analysis and the construction of histograms is available as Supporting Information.

# Supporting Information

**Conductance Histogram based on all-data point analysis**

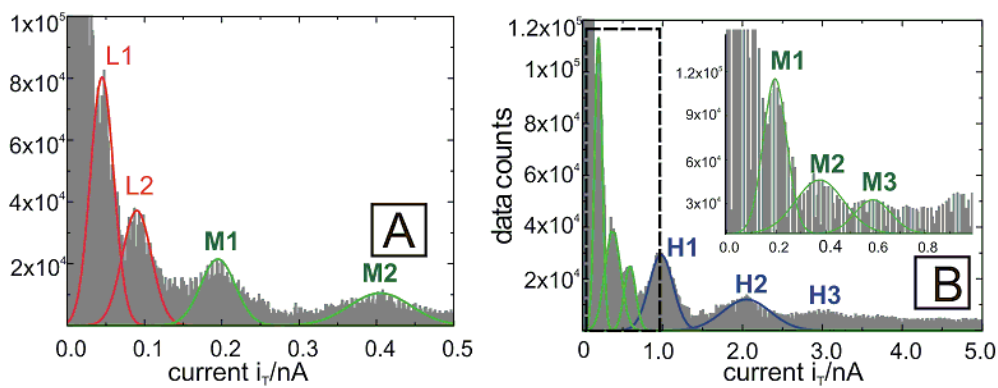

***Figure S-1:*** All data point conductance histogram constructed from type III "stretching" curves for Au│1,9-nonanedithiol│Au junctions. (A) 1600 out of 4300 traces employing the 1 nA (max) preamplifier; (B) 1100 out of 4300 traces recorded with the 10 nA (max) preamplifier. All other conditions are identical to those in Fig. 3. The inset in (B) shows the magnified current region below 1 nA. The solid red, green and blue lines illustrate up to three peaks of the H, M and L conductance junctions. Compared to the analysis of plateau currents, the all-data point conductance histograms exhibit a low current region, which is less clearly resolved as well as a larger "offset". However, the single junction conductance data as extracted for each sequence are in excellent agreement with the results obtained from the analysis of plateau currents (in brackets): L: 0.45 nS (0.47 nS); M: 1.95 nS (2.0 nS) and H: 9.8 nS (9.9 nS). For a comprehensive and comparative discussion of various strategies of constructing and analyzing conductance histograms with refer to a forthcoming paper (Ref. 38).

**Computational details for constructing the histograms in Fig. 8B (Ref. 59)**

Conductance histograms for *n*-alkanedithiols, for example like the one shown in Fig. 8B where *n*=9, were simulated by summing up appropriately weighted ($w_\mu$) Gaussian distributions $P(g) = \sum_\mu \frac{w_\mu}{\sqrt{2\pi\sigma_\mu^2}} \exp\left[-(g - g_\mu)^2 / 2\sigma_\mu^2\right]$ ascribed to each conductance value $g_\mu$, which a molecular bridge, consisting in general of more than just a single molecule, could have. Here $\mu=(m; k\alpha, k'\alpha', k''\alpha'', …)$ is a multi-index labeling the *m*-molecule junction, conformations $k, k', k''…$ (trans/gauche) of each molecule in the bridge, and types $\alpha, \alpha', \alpha''$, ("atop-atop", "atop-bridge", etc …) of the electrodes-molecule coupling. Let $k = 0$ denote the straight (trans) isomers. For the gauche-isomers with a single defect, $k$ could denote a C-C bond over which rotation is performed. In case of more than one defect, $k$ itself is obviously a multi-index. Conductance histograms were simulated for the range of low (L), medium (M) and H (high) values and were summed up assuming arbitrary relative weights (integrated areas).

In order to reach the smallest possible conductance values $g$ for the L-series, only "atop-atop" contact geometries (fixed $\alpha$) and only gauche isomers were counted. We considered calculated conductance values $g_k$ due to single molecules ($m=1$) with gauche defects (weights $w_k = 1$), as well as pairs of molecules ($m=2$) with conductances $g_{kk'} = g_k + g_{k'}$ and some statistical weights $w_{kk'} = p(n)$, and triplets of molecules ($m=3$) with conductances $g_{kk'k''} = g_k + g_{k'} + g_{k''}$ and $w_{kk'k''} = p^2(n)$. With an eye on experimental data we found $p(n=9) = 0.2$ as a reasonable choice in case of 1,9-nonanedithiol. For molecules with another lengths $n$, the pair weights $p(n)$ were taken as $p(n) = p(9)N(n)/N(9)$ according to the number $N(n)$ of gauche conformations for a given $n$, which was 10, 15, 6, 7, and 8 for $n = 5, 6, 7, 8,$ and 9, respectively. To improve statistics, for short species with $n = 5$ and 6 we took into account molecules with one and two structural defects, while only single defects were counted for longer ($n=7,8,9$) molecules. Everywhere, we assumed $\sigma_m \sim \sqrt{g_m}$ for the standard deviations.

A similar procedure was employed for the M and H-series. We took only two calculated conductance values for "atop-atop" and "bridge-bridge" coordination of molecules in an all-trans conformation ($k=0$), and assumed that bridges with up to $m=4$ molecules could appear. The conditional weights $w_m=(p')^{m-1}$ were powers of $p'$. The value $p'$ is somewhat arbitrary parameter, which should be chosen between $p(n)$ and 1.0; a reasonable choice was $p' = 0.75$. The contribution of other configurations to the histogram (with atomic chains bridged between adatoms, as well as with "bridge-atop" or "hollow-hollow" coordination) would lead only to increased weights of already existing peaks. Such configurations show conductances either very close to the medium values or twice as high, and therefore have been ignored. Due to the observed tendency for different gauche conformations (for a given $n$) to show conductances close to multiples of some smaller "master" value, we were able to construct conductance histograms with up to four pronounced, equidistantly separated peaks within each set of L, M, or H values for all $n$-alkanedithiols studied experimentally.